\documentclass[12pt]{iopart}

\usepackage{graphicx}
\usepackage{iopams}
\usepackage{amssymb}
\usepackage{hyperref}
\usepackage{booktabs}
\usepackage{xcolor}
\usepackage{tikz}

\newcommand{\binom}[2]{\Bigl(\begin{array}{@{}c@{}}#1\\#2\end{array}\Bigr)}

\newcommand{\BigO}[1]{\mathcal{O}(#1)}
\newcommand{\Binomial}{\mathrm{Bin}}
\newcommand{\Mult}{\mathrm{Mult}}

\newcommand{\seqbin}{\textsf{SeqBin}}

\newcounter{algorithm}
\newcommand{\algcaption}[1]{%
  \refstepcounter{algorithm}%
  \noindent\textbf{Algorithm~\thealgorithm:} #1}

\begin{document}

\title{Multinomial shot sampling for quantum circuits}

\author{Ijaz Ahamed Mohammad$^1$, Daniel Nagaj$^{1}$ and Martin Plesch$^{1,2}$}

\address{$^1$ Institute of Physics, Slovak Academy of Sciences, D\'ubravsk\'a cesta 9, 845~11 Bratislava, Slovakia}
\address{$^2$ Matej Bel University, N\'arodn\'a ulica 12, 974~01 Bansk\'a    Bystrica, Slovakia}
\ead{ijaz.ahamed@savba.sk} 

\begin{abstract}
In classical simulations of quantum circuits, sampling shots (measurement
outcomes) from the final state is a routine but costly step. In widely used
software, including Qiskit and PennyLane, its cost grows linearly with the
number of shots~$S$, even when the simulation has already produced the full
state vector, because every shot is generated and processed individually. If
only the measurement counts are needed, this is unnecessary. The counts can be
drawn directly from a multinomial distribution, decomposed into a sequence of
conditional binomial draws, each performed in expected constant time by the
BTPE algorithm. The sampling cost is then independent of~$S$. We describe an implementation of this approach, which also
shortens the computation for states whose probability is concentrated on a few
outcomes, and compare it with Qiskit's sampling primitives for up to $20$
qubits and $10^6$ shots. Our implementation is faster in every
configuration tested, by a factor that grows with the number of shots and
exceeds $10^3$ in most configurations at $S = 10^6$, and its samples are
statistically indistinguishable from those of the reference implementations.
%
\end{abstract}

\noindent{\it Keywords\/}: quantum simulation, shot sampling, multinomial sampling, binomial distribution, BTPE, variational quantum eigensolver


\section{Introduction}
\label{sec:intro}

Current quantum computers are still limited by noise, scale, and hardware availability. As a result, classical simulation remains an important tool for developing and testing quantum algorithms before they can be reliably executed on sufficiently capable quantum hardware. A classical simulation can reproduce not only the evolution of a quantum state, but also its measurement process. Once the final quantum state is known, it defines a probability distribution over all possible measurement outcomes. To imitate measurements on a quantum computer, outcomes are then sampled from this distribution according to a chosen number of shots, i.e., repeated quantum measurements.

For a small number of shots, the choice of sampling method is usually not important. Each shot can simply be generated independently from the probability distribution. However, when the number of shots becomes very large, repeatedly generating and storing individual outcomes can itself become computationally expensive. This is particularly relevant when the same simulated state must be sampled many times. A natural question is therefore whether every shot needs to be simulated individually.

If only the final measurement counts are required, this is not necessary. Instead of generating the outcomes one shot at a time and counting them afterwards, the complete set of outcome counts can be drawn directly from a multinomial distribution. This becomes particularly attractive when the number of shots is large compared with the number of possible measurement outcomes. In this regime, the sampling procedure can be organized around the number of possible outcomes rather than around the total number of shots. The task is then no longer to generate every measurement outcome individually, but to draw a single multinomial sample, that is, to produce the entire vector of outcome counts in one step. This is what we refer to as multinomial sampling.

This is a well established problem in the statistical simulation literature. The standard count-level construction is a sequence of conditional binomial draws~\cite{Devroye1986}, with the binomial variates supplied efficiently by, for example, the BTPE algorithm of Kachitvichyanukul and Schmeiser~\cite{BTPE1988,Cicirello2024}.

 A simulator that requires only the measurement counts would therefore be expected to use this construction. The current Qiskit~\cite{Qiskit} reference workflow nevertheless remains shot-resolved: individual measurement outcomes are generated and subsequently converted into the requested result representation. The sampling work therefore continues to scale with the number of shots, even though the quantum state and its measurement probabilities are already available. This distinction has also been discussed within the Qiskit community~\cite{QiskitIssue8618,NoteQiskitDiscussion}. The same shot-resolved design is used in PennyLane~\cite{PennyLane}, so two of the most widely used quantum software development kits share this cost. We study the Qiskit workflow in detail and describe the PennyLane path briefly in \ref{app:qiskit}.

 In this work we show that count-level sampling substantially reduces the runtime of the shot-sampling step in classical quantum-circuit simulation. We determine where the current reference workflow spends its time, present an implementation that exploits the structure of quantum-state probability distributions, and verify that the resulting samples remain statistically exact.

The remainder of the paper is organized as follows. Section~\ref{sec:problem} introduces the shot-sampling problem and briefly describes the sampling approaches commonly used in classical quantum-circuit simulation. Section~\ref{sec:method} presents the multinomial sampling method based on sequential conditional binomial draws and summarizes the practical modifications used in our implementation. Section~\ref{sec:comparison} benchmarks the different sampling approaches across a range of shot counts, system sizes, and quantum-state distributions, and validates that they produce statistically equivalent results. Section~\ref{sec:conclusion} summarizes the regimes in which multinomial sampling is most useful and its relevance to current quantum-simulation workflows. Additional algorithmic details and technical derivations are provided in the Appendix.

\section{Problem setup}
\label{sec:problem}

Consider a quantum circuit on $n$ qubits. After classical simulation, the final quantum state defines a probability distribution

\begin{equation}
    \mathbf{p} = (p_1,p_2,\ldots,p_D),
    \qquad D = 2^n,
\end{equation}

over the possible measurement outcomes, where

\begin{equation}
    \sum_{i=1}^{D} p_i = 1.
\end{equation}

For a simulation with $S$ shots, the objective is to generate measurement counts that are consistent with this probability distribution. Here, we are not concerned with the cost of obtaining the quantum state itself, but with the subsequent sampling step once the probability distribution is already available.

\subsection{Shot-by-shot sampling}
\label{sec:shotbyshot}

A straightforward approach is to generate each measurement outcome independently. This can be done using the cumulative distribution function (CDF),

\begin{equation}
    C_i = \sum_{j=1}^{i} p_j.
\end{equation}

For each shot, a uniformly distributed random number $u\in[0,1)$ is generated, and the corresponding outcome is found by locating the first cumulative probability $C_i$ that exceeds $u$. Drawing a single outcome from a fixed list of probabilities in this way is called a categorical draw, and repeating it once per shot is what we refer to as categorical sampling.

Since the CDF is an ordered array containing $D$ values, this lookup can be performed using binary search. At each comparison, approximately half of the remaining possible outcomes can be discarded: from $D$ possibilities to $D/2$, then $D/4$, and so on. After approximately $\log_2 D$ comparisons, only one outcome remains. The lookup therefore requires $\BigO{\log D}$ operations for each shot. Since $D=2^n$, this is equivalent to $\BigO{n}$ operations per shot.

Repeating the procedure independently for $S$ shots gives a sampling cost of $\BigO{S\log D}$. In addition, the CDF itself must first be constructed from the $D$ probabilities, which requires $\BigO{D}$ operations. The total cost is therefore $\BigO{D + S\log D}$, or, using $D=2^n$, $\BigO{2^n + Sn}$.

Thus, although generating a single measurement outcome is relatively inexpensive, the search must be repeated for every shot. Consequently, the sampling time continues to increase with $S$, even after the quantum state and its probability distribution have already been computed (\ref{app:qiskit}).

\subsection{Sampling the counts directly}

If only the final measurement counts are required, generating and storing the individual shot outcomes is unnecessary. The complete count vector

\begin{equation}
    \mathbf{k} = (k_1,k_2,\ldots,k_D),
\end{equation}

where

\begin{equation}
    \sum_{i=1}^{D} k_i = S,
\end{equation}

follows a multinomial distribution,

\begin{equation}
    \mathbf{k} \sim
    \Mult\!\left(S;\mathbf{p}\right).
\end{equation}

The sampling problem can therefore be reformulated from generating $S$ individual outcomes to generating the $D$ outcome counts directly.

A multinomial draw can be generated through a sequence of conditional binomial draws. If each binomial random variable can be generated in expected constant time, the complete multinomial sample requires an expected time of $\BigO{D} = \BigO{2^n}$, with no direct dependence on the total number of shots $S$. The main difference between the two approaches can therefore be summarized as shot-by-shot sampling:  $\BigO{D + S\log D}$, compared with count-level multinomial sampling: $\BigO{D}$.

 A sampling run is informative only when the number of shots is sufficient to populate the outcomes that carry the probability. For a distribution spread over the full outcome space this requires $S \gtrsim D$; for a concentrated distribution it requires $S$ to exceed the size of the support. These are the regimes in which measurement data is collected in practice, and they are the regimes considered here. Throughout these regimes the count-level approach is the faster of the two, by a margin that increases with the shot budget. Section~\ref{sec:comparison} presents the comparison in detail~\cite{NoteLowShotRegime}.

\subsection{Inefficiencies in current simulation tools}

The distinction between shot-by-shot and count-level sampling is also relevant to current quantum-simulation software. In Qiskit's reference
\texttt{StatevectorSampler}~\cite{QiskitSampler}, the probability distribution is first obtained from the simulated state, after which the requested measurement outcomes are generated individually for all $S$ shots. These individual outcomes must then be stored and converted into the result representation returned to the user. Thus, even after the quantum state and its measurement probabilities are already available, additional work is performed for every requested shot.

Our benchmarks show that this per-shot processing can become considerably more expensive than the statevector simulation itself when $S$ is large, and that the $\BigO{S}$ sampling step accounts for only part of the cost. Circuit preparation and transpilation, backend initialisation, primitive dispatch, and in particular the materialization, conversion and packaging of shot-resolved results all contribute, and each is incurred on every call rather than once. Where possible, one-time initialisation and compilation costs are removed from the timed measurements, so that the comparison reported here is favourable to the existing tools. The Aer-based sampler follows a different, compiled sampling path, but its runtime also retains a strong dependence on the number of shots. \ref{app:qiskit} gives the stage-by-stage breakdown.

\section{Method: the \seqbin{} sampler}
\label{sec:method}

 The count-level sampler used in this work, which we denote \seqbin,
generates the multinomial count vector from a sequence of conditional
binomial draws. It consists of two components: an efficient primitive for
generating a single binomial variate, and a decomposition that combines
these variates into one multinomial sample. The two components are
introduced in that order, followed by the practical refinements used in the
benchmarks.

\subsection{The BTPE algorithm}
\label{sec:btpe}

The multinomial sampler described in the next section is built entirely from
repeated binomial draws, so we first need an efficient way to generate a single
one. Let
\begin{equation}
    X \sim \Binomial(N,p)
\end{equation}
be the number of successes in $N$ independent trials, each succeeding with
probability $p$. The obvious way to obtain $X$ is to simulate the $N$ trials and
count the successes. This, however, reproduces exactly the difficulty we are
trying to avoid: the work grows with $N$, even though only a single number is
needed at the end.

The BTPE algorithm of Kachitvichyanukul and Schmeiser~\cite{BTPE1988,BTPEC1989} obtains
$X$ without ever touching the individual trials. It is built on
acceptance--rejection sampling, a general technique for drawing from a
distribution that is inconvenient to sample from directly. Picture the binomial
distribution as a bar chart, with a bar of height $P(X=k)$ standing at every
possible count $k$. Sampling from the distribution is then equivalent to picking
a point at random from somewhere inside those bars and reporting the count $k$
that the point stands on, since a taller bar is hit more often than a short one.
The bars themselves are awkward to sample from, so a simpler shape is placed
over them, high enough to cover every bar. A candidate point is drawn uniformly
from beneath that simpler shape, which is easy by construction. If the candidate
lands inside a bar it is kept; if it lands in the empty gap between the top of a
bar and the shape above it, the candidate is discarded and a new one is drawn.
Every point that survives this test lies inside a bar, and the survivors are
spread evenly across the bars. A bar twice as tall therefore keeps twice as
many of them, so each count $k$ comes out with probability proportional to
$P(X=k)$, which is exactly the binomial distribution. Note that the covering
shape changes how many candidates are thrown away before one is kept, but not
which values are kept; the result is exact, not an approximation.

\begin{figure}[t]
\centering
\resizebox{0.98\linewidth}{!}{%
\begin{tikzpicture}[
    x=0.55cm,
    y=0.72cm,
    line cap=round,
    line join=round,
    font=\scriptsize
]
\begin{scope}
    \fill[orange!70]
        (0.25,0) -- (0.25,0.5608)
        -- plot[domain=0.25:2.15,samples=40]
            (\x,{1.45*exp(0.50*(\x-2.15))})
        -- (2.15,0) -- cycle;
    \fill[teal!22]
        (2.15,0) -- (2.15,1.45) -- (3.85,2.65)
        -- (3.85,0) -- cycle;
    \fill[blue!18]
        (3.85,0) -- (3.85,2.65) -- (5.15,4.0)
        -- (6.45,3.3) -- (6.45,0) -- cycle;
    \fill[violet!22]
        (6.45,0) -- (6.45,3.3) -- (8.15,2.8)
        -- (8.15,0) -- cycle;
    \fill[orange!70]
        (8.15,0) -- (8.15,2.8)
        -- plot[domain=8.15:10.75,samples=40]
            (\x,{2.8*exp(-0.42*(\x-8.15))})
        -- (10.75,0) -- cycle;

    \draw[->] (0,0) -- (11.5,0) node[below=2pt] {$k$};
    \draw[->] (0,0) -- (0,4.0) node[left=2pt] {$P(X=k)$};

    \foreach \x/\h in {
        0.5/0.18, 1.5/0.34, 2.5/0.65, 3.5/1.10, 4.5/1.80,
        5.5/2.45, 6.5/2.80, 7.5/2.35, 8.5/1.60, 9.5/0.85, 10.5/0.40}
        {\draw[fill=gray!45,fill opacity=0.48,draw=gray!70!black]
            (\x-0.27,0) rectangle (\x+0.27,\h);}

    \fill[orange!70,fill opacity=0.45]
        (0.25,0) -- (0.25,0.5608)
        -- plot[domain=0.25:2.15,samples=40]
            (\x,{1.45*exp(0.50*(\x-2.15))})
        -- (2.15,0) -- cycle;
    \fill[orange!70,fill opacity=0.45]
        (8.15,0) -- (8.15,2.8)
        -- plot[domain=8.15:10.75,samples=40]
            (\x,{2.8*exp(-0.42*(\x-8.15))})
        -- (10.75,0) -- cycle;

    \draw[blue!75!black, very thick]
        plot[domain=0.25:2.15,samples=40]
            (\x,{1.45*exp(0.50*(\x-2.15))})
        -- (3.85,2.65) -- (5.15,4.0) -- (6.45,3.3)
        -- (8.15,2.8)
        -- plot[domain=8.15:10.75,samples=40]
            (\x,{2.8*exp(-0.42*(\x-8.15))});

    \draw[densely dashed,gray!70!black]
        (2.15,0) -- (2.15,1.45)
        (3.85,0) -- (3.85,2.65)
        (6.45,0) -- (6.45,3.3)
        (8.15,0) -- (8.15,2.8);

    \node[fill=orange!70,inner sep=1pt] at (1.15,-0.34) {Exp. tail};
    \node[fill=teal!22,inner sep=1pt] at (3.0,-0.34) {Para.};
    \node[fill=blue!18,inner sep=1pt] at (5.15,-0.34) {Triangle};
    \node[fill=violet!22,inner sep=1pt] at (7.3,-0.34) {Para.};
    \node[fill=orange!70,inner sep=1pt] at (9.1,-0.34) {Exp. tail};
    \node[font=\footnotesize] at (5.75,4.38) {(a) Proposal envelope};
\end{scope}

\begin{scope}[xshift=8cm]
    \draw[->] (0,0) -- (11.5,0) node[below=2pt] {$k$};
    \draw[->] (0,0) -- (0,4.0) node[left=2pt] {height};

    \draw[fill=gray!45,draw=gray!70!black]
        (3.45,0) rectangle (4.55,2.4);
    \draw[dashed,gray!70!black] (4,0) -- (4,3.35);
    \draw[dashed,gray!60!black] (0.8,2.4) -- (10.6,2.4);
    \draw[blue!75!black, very thick] (0.8,3.35) -- (10.6,3.35);

    \fill[green!55!black] (4,1.55) circle (0.11);
    \fill[red!70!black] (4,2.85) circle (0.11);
    \node[anchor=west] at (4.45,1.55) {accept: $y\leq P(X=k)$};
    \node[anchor=west] at (4.45,2.85) {reject: $y>P(X=k)$};
    \node[anchor=west] at (10.65,3.35) {$h(k)$};
    \node[anchor=west] at (10.65,2.4) {$P(X=k)$};
    \node[font=\footnotesize] at (5.75,4.38) {(b) Accept--reject test};
\end{scope}
\end{tikzpicture}
}%
\caption{Schematic illustration of the BTPE algorithm. The grey bars represent
the exact binomial probabilities, while the blue line represents the proposal
envelope. In panel (a), the envelope is divided into one central triangular
region, one parallelogram on each side, and two outer exponential tails; the
colours and labels distinguish these five regions.
In panel (b), a candidate point $(k,y)$ is sampled under the envelope. At the
selected value of $k$, the candidate is accepted if $y\leq P(X=k)$ and rejected
otherwise.}
\label{fig:btpe}
\end{figure}
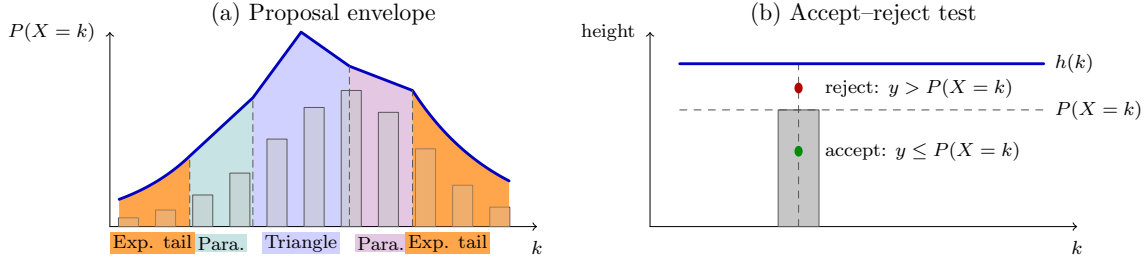

The covering shape is called the proposal envelope, and everything therefore
depends on choosing one that is both easy to sample from and close to the
binomial probability mass function. BTPE takes its name from the pieces it uses:
a central \emph{triangle} over the bulk of the distribution, a
\emph{parallelogram} on either side of it, and an \emph{exponential} tail at
each end, together covering the \emph{binomial} mass. This mirrors the shape of
the binomial distribution itself, which is peaked in the middle and decays
quickly on both sides, and each of the five pieces can be sampled with a few
arithmetic operations. Figure~\ref{fig:btpe} shows the construction.

A single attempt therefore selects one of the five regions, draws a candidate
pair $(k,y)$ from it, and tests whether the height $y$ falls inside the bar at
$k$, that is whether $y \leq P(X=k)$. Evaluating $P(X=k)$ costs a fixed amount
of work that does not depend on $N$, as described in \ref{app:btpe}. Because the
envelope follows the distribution closely, few candidates land in the gap, and
the expected number of attempts per draw stays small and does not grow with
$N$~\cite{Cicirello2024}. This is the property that the rest of the method rests on: one binomial draw has
an effectively constant expected cost, independent of the number of trials $N$.

The envelope construction requires the distribution to be broad enough to have a
well-defined bulk and tails, which in practice means that the expected numbers of
successes and failures are both large, typically $Np \geq 30$ and
 $N(1-p) \geq 30$. Outside this regime the envelope construction is no
longer efficient, and inversion or another direct small-mean method is both
faster and simpler. The binomial distribution being sampled is unchanged, so
the choice between them is a matter of method selection rather than of
approximation or loss of correctness. Standard implementations therefore
switch between BTPE and a direct method according to the values of $Np$ and
$N(1-p)$. The remaining technical details are collected in \ref{app:btpe}.

\subsection{Multinomial sampling via conditional binomials}
\label{sec:multinomial}

 The key observation is that a sample from a $D$-outcome multinomial
distribution can be generated by a sequence
of $D-1$ conditional binomial draws. This is a standard technique in the random
variate generation literature~\cite{Devroye1986}. Given the probability vector
$\mathbf{p} = (p_1, \ldots, p_D)$ with $D = 2^n$ and a total of $S$ shots, the
first count is drawn as
\begin{equation}
    k_1 \sim \Binomial(S,p_1).
\end{equation}
After this draw, $S-k_1$ shots remain to be assigned among the remaining
outcomes, whose total probability is $1-p_1$. The next count is therefore drawn
conditionally as
\begin{equation}
    k_i \sim \Binomial\!\left(S_\mathrm{rem},
    \frac{p_i}{p_\mathrm{rem}}\right),
\end{equation}
where $S_\mathrm{rem}$ is the number of unassigned shots and
$p_\mathrm{rem}$ is the total probability of the unprocessed outcomes. Repeating
this operation for $i=1,\ldots,D-1$ leaves the final count fixed by
\begin{equation}
    k_D = S_\mathrm{rem}.
\end{equation}
The resulting vector satisfies
$\mathbf{k} \sim \Mult(S;\mathbf{p})$.

The procedure is summarized in Algorithm~\ref{alg:seqbinom}.

\medskip
\algcaption{\seqbin: conditional-binomial decomposition for multinomial sampling.}\label{alg:seqbinom}\\[4pt]
\noindent\textbf{Input:} Probability vector $\mathbf{p} = (p_1, \ldots, p_D)$, shot count $S$.\\
\noindent\textbf{Output:} Count vector $\mathbf{k} = (k_1, \ldots, k_D)$ with $\mathbf{k} \sim \Mult(S; \mathbf{p})$.
\begin{quote}
\begin{tabbing}
\hspace{4mm}\=\hspace{4mm}\=\hspace{4mm}\=\hspace{4mm}\=\hspace{4mm}\=\kill
1.\>\> $S_\mathrm{rem} \gets S$\\
2.\>\> $p_\mathrm{rem} \gets 1$\\
3.\>\> \textbf{for} $i = 1, \ldots, D-1$ \textbf{do}\\
4.\>\>\> \textbf{if} $S_\mathrm{rem} = 0$ \textbf{then}\\
5.\>\>\>\> $k_i \gets 0$ \hfill\textrm{(early stopping)}\\
6.\>\>\> \textbf{else}\\
7.\>\>\>\> $k_i \gets \mathrm{BinomialDraw}(S_\mathrm{rem},\; p_i / p_\mathrm{rem})$\\
8.\>\>\>\> $S_\mathrm{rem} \gets S_\mathrm{rem} - k_i$\\
9.\>\>\>\> $p_\mathrm{rem} \gets p_\mathrm{rem} - p_i$\\
10.\>\>\> \textbf{end if}\\
11.\>\> \textbf{end for}\\
12.\>\> $k_D \gets S_\mathrm{rem}$
\end{tabbing}
\end{quote}

Each binomial draw in this procedure can use BTPE and therefore
has effectively constant expected cost. The complete decomposition then
requires an expected $\BigO{D}=\BigO{2^n}$ operations, independent of the total
number of shots $S$, apart from the cost of reading and preparing the probability
vector.

\subsection{Practical optimizations}
\label{sec:optimizations}

 Several refinements improve both the speed and numerical stability of
this procedure.

\subsubsection{Bunching rare outcomes.}
\label{sec:bunching}
For large~$n$, most outcomes carry very little probability. If $p_i \ll 1/S$
then the expected count $S p_i$ is much smaller than one, so the binomial draw
for that outcome almost always returns zero. Spending a draw on each of them
individually is wasteful: the sweep does $D$ units of work to discover that
nearly all of the answers are zero.

Instead we split the outcomes into a ``common'' set ($p_i > \theta$) and a
``rare'' set ($p_i \leq \theta$), using a threshold $\theta$ that marks the
point below which an outcome is not expected to receive even one shot
(e.g.\ $\theta = c/S$ for a small constant~$c$). The rare outcomes are then
handled together rather than one at a time:
\begin{enumerate}
    \item Perform the \seqbin{} decomposition over the common
    outcomes and a single aggregate rare bin with probability
    $p_\mathrm{rare} = \sum_\mathrm{rare} p_i$.
    \item Place the aggregate bin last in the sweep, so that its total count
    $k_\mathrm{rare}$ is the number of shots left after the common outcomes
    have been drawn.
    \item If $k_\mathrm{rare} > 0$, distribute those shots among the individual
    rare outcomes by categorical sampling, drawing them one at a time from the
    rare probabilities alone. This is the shot-by-shot method of
    Section~\ref{sec:shotbyshot}, and it is cheap here precisely because
    $k_\mathrm{rare}$ is small: it is the same trade-off made in
    \ref{sec:switching}.
\end{enumerate}
 This reduces the number of conditional binomial draws from~$D-1$ to at
most $D_\mathrm{common}$, where $D_\mathrm{common}$ is the number of common
outcomes. The saving can be substantial when
the distribution is spread across many low-probability outcomes.

\subsubsection{Minor refinements.}
\label{sec:implementation_details}
The implementation includes three further refinements, which affect
robustness rather than the dominant cost: the remaining probability is
obtained from precomputed suffix sums rather than from a running
subtraction, the sweep terminates once every shot has been assigned, and a
short unassigned tail is completed by categorical sampling. The
rare-outcome threshold additionally requires a safeguard, since a fixed
choice degrades as the shot count approaches the number of outcomes.
Finally, the outcomes are visited in their natural order rather than sorted by
probability, since a sort would cost more than the sweep it is meant to
accelerate. These points are described in \ref{app:implementation}.

\subsection{Complexity summary}
\label{sec:complexity}

 The costs of the two approaches can now be summarized. Shot-by-shot
categorical sampling requires $\BigO{D + S\log D}$ operations, since the
cumulative distribution is constructed once and one binary search is then
performed per shot. The \seqbin{} sweep requires $\BigO{D}$ operations,
since each of at most $D-1$ conditional draws takes expected constant time;
with rare-outcome bunching the sweep runs over $D_\mathrm{common} + 1$
effective bins rather than over all $D$. Neither of the count-level costs
depends on $S$.
NumPy's multinomial generator uses the same conditional-binomial principle
internally~\cite{NumpySource} and shares the $\BigO{D}$ scaling; the comparison
in this paper therefore separates this count-level family as a whole from
Qiskit's shot-resolved workflow.

\section{Experimental results and comparison}
\label{sec:comparison}

\subsection{Benchmarking setup}
\label{sec:bench_setup}

\subsubsection{Count-level methods operating on the probability vector.}
These methods take the probability vector as given and return a count
vector. We compare NumPy's compiled multinomial generator with three
implementations of \seqbin{} that differ only in how each binomial variate
is generated:
\begin{enumerate}
    \item \textbf{NumPy multinomial}: a direct call to
    \texttt{numpy.random.multinomial} with the precomputed probability vector.
    \item \textbf{\seqbin--SciPy}: Algorithm~\ref{alg:seqbinom} using
    \texttt{scipy.stats.binom.rvs} for each draw.
    \item \textbf{\seqbin--Python}: the same decomposition with a direct BTPE
    implementation in pure Python, avoiding SciPy's internal method-switching
    overhead.
    \item \textbf{\seqbin--Numba}: the same implementation again, with its
    inner loop translated into machine code before it runs, using Numba. This
    removes the interpreter overhead of the pure Python version and puts the
    comparison against NumPy's own compiled loop on an equal footing.
\end{enumerate}

\subsubsection{Shot-resolved Qiskit workflows.}
These methods take a circuit as input and return a shot-resolved result
object, and therefore perform more work than a count draw alone. They are
included because they are the user-facing workflows whose measurement
overhead motivates this study:
\begin{enumerate}
    \item \textbf{Qiskit StatevectorSampler}: the current Qiskit reference
    primitive, and the only built-in sampler in \texttt{qiskit.primitives}~\cite{Qiskit20ReleaseNotes}.
    \item \textbf{Qiskit Aer SamplerV2}: the V2 Aer-backed primitive
    (\texttt{qiskit\textunderscore \linebreak[0]aer.primitives.\linebreak[0]SamplerV2})~\cite{QiskitAerDocs}.
\end{enumerate}

\subsubsection{Benchmark grid and timing protocol.}

\noindent
All benchmarks were performed on the Perun HPC system at the Technical University
of Košice~\cite{perun}. The software environment was Python 3.9.25,
NumPy 2.0.2~\cite{NumPy}, SciPy 1.13.1~\cite{SciPy}, Qiskit 2.2.3~\cite{Qiskit},
Qiskit-Aer 0.17.2~\cite{QiskitAerDocs}, and Numba 0.60.0~\cite{Numba}.
For each configuration, we report wall-clock runtimes averaged over 50 independent
runs for the probability-vector methods and 10 independent runs for the Qiskit
sampler primitives. For the Qiskit primitives each timed call runs from the
circuit to the counts, including their extraction from the returned
\texttt{BitArray} into a count vector, so that every method returns the same
output. Each job was allocated four CPU cores (eight for $n \ge 16$); NumPy
multinomial and the \seqbin{} implementations run on a single core, whereas Aer
can use several.

We test across:
\begin{itemize}
    \item \textbf{System sizes:} $n = 2, 4, 6, \ldots, 20$ qubits (i.e.\ $D = 4$ to $D \approx 10^6$).
    \item \textbf{Shot counts:} $S = 10^4, 10^5, 10^6$.
    \item \textbf{State types:}
    (a)~random circuits of depth~10, in which each layer applies $R_Y$ and $R_Z$
    rotations with uniformly random angles to every qubit, followed by a chain
    of CNOT gates between neighbouring qubits (generic probability distributions),
    (b)~equal superposition $p_i = 2^{-n}$ (uniform),
    (c)~GHZ states (maximally sparse: two nonzero outcomes),
    (d)~generalized W~states (sparse: $n$ nonzero outcomes with equal probability).
\end{itemize}

All figures report the mean runtime with error bars indicating one standard deviation. For the fastest methods at small $n$, where a call takes only microseconds, the run-to-run spread is comparable to the mean, which shows as long error bars on the logarithmic axes.

Before timing each Qiskit sampler, we performed two un-timed warmup calls using
the same circuit and the same shot count as the subsequent timed calls, and discarded
their outputs. These warmup calls remove one-time costs such as sampler construction,
backend initialisation, transpilation and cache setup, and one-time compilation, from the
reported steady-state timings. The reported Qiskit runtimes should therefore be
interpreted as favourable steady-state per-call costs rather than first-call user latency.
For \seqbin--Numba, we similarly performed one un-timed dummy call on
a small probability vector, so that the one-time cost of compiling it is excluded from the reported timings.

\vspace{5mm}

\subsection{Scaling with the number of shots}
\label{sec:scaling_shots}

\begin{figure}[h]
    \centering
    \includegraphics[width=\linewidth]{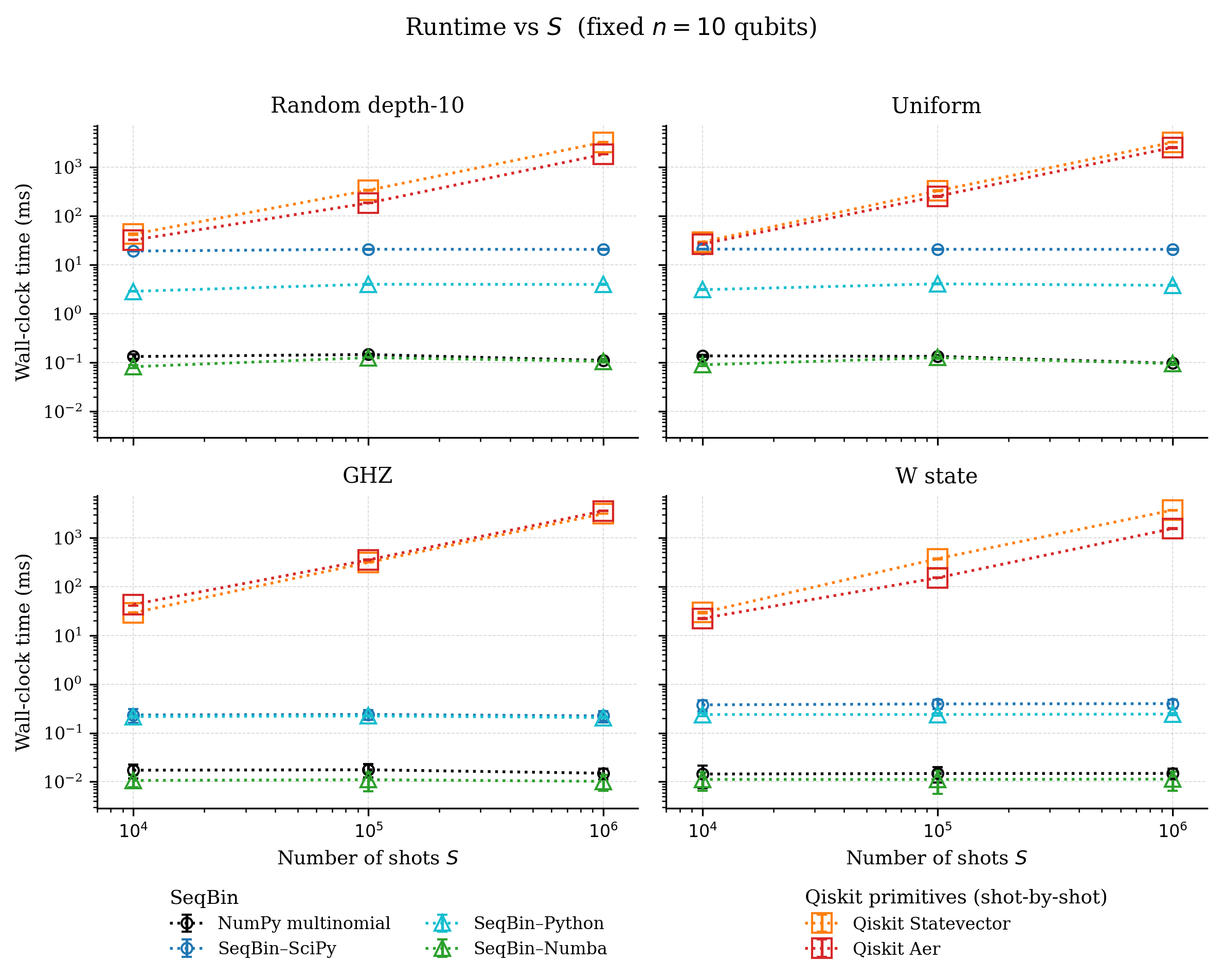}
    \caption{
		Wall-clock runtime as a function of the shot budget $S$ at $n = 10$ qubits, with one panel per state type (Random depth-10, Uniform, GHZ, W). Each marker is the mean over 50 independent repeats for the count-level methods (NumPy multinomial and the three \seqbin{} backends) and 10 repeats for the Qiskit primitives; error bars show one standard deviation and are visible mainly for the fastest methods on the sparse states. \seqbin--Numba was run once on a small input before timing began, and each Qiskit sampler was instantiated and run twice before the first timed call, so the reported numbers exclude one-time compilation and transpilation costs.}
    \label{fig:runtime vs S}
\end{figure}

Figure~\ref{fig:runtime vs S} shows runtime as a function of shot count $S \in {10^4, 10^5, 10^6}$ for fixed $n = 10$ qubits across all four state types. The two Qiskit methods, StatevectorSampler and Aer Sampler, scale linearly with $S$. By contrast, all four count-level methods remain flat across the entire range. NumPy multinomial and \seqbin--Numba sit near $0.1$ ms on dense states. \seqbin--Python pays a constant interpreter penalty, sitting about thirty times higher. \seqbin--SciPy lies higher still on dense states, about $20$ ms. On the sparse states only a few outcomes carry probability mass: two for GHZ, independently of $n$, and $n$ for the W state, which is ten here. Rare-outcome bunching collapses the sweep to those few bins, and \seqbin--Numba falls to about $0.01$ ms, below NumPy multinomial, which visits all $D = 1024$ outcomes and takes about $0.015$ ms. Neither Qiskit primitive benefits from state sparsity. \texttt{StatevectorSampler} draws each shot independently with \texttt{numpy.random.choice}, so its cost is set by $S$ alone and is flat across the four state types to within about twenty per cent for $S \geq 10^5$. Aer uses a different, compiled sample-measure path, and its runtime does vary by about a factor of two between state types; that variation does not follow the support size, however, since GHZ, the sparsest of the four, is the slowest. It reflects the cost of simulating the different circuits rather than the number of outcomes carrying probability.


\subsection{Scaling with system size}
\label{sec:scaling_n}

\begin{figure}[h]
    \centering
    \includegraphics[width=\linewidth]{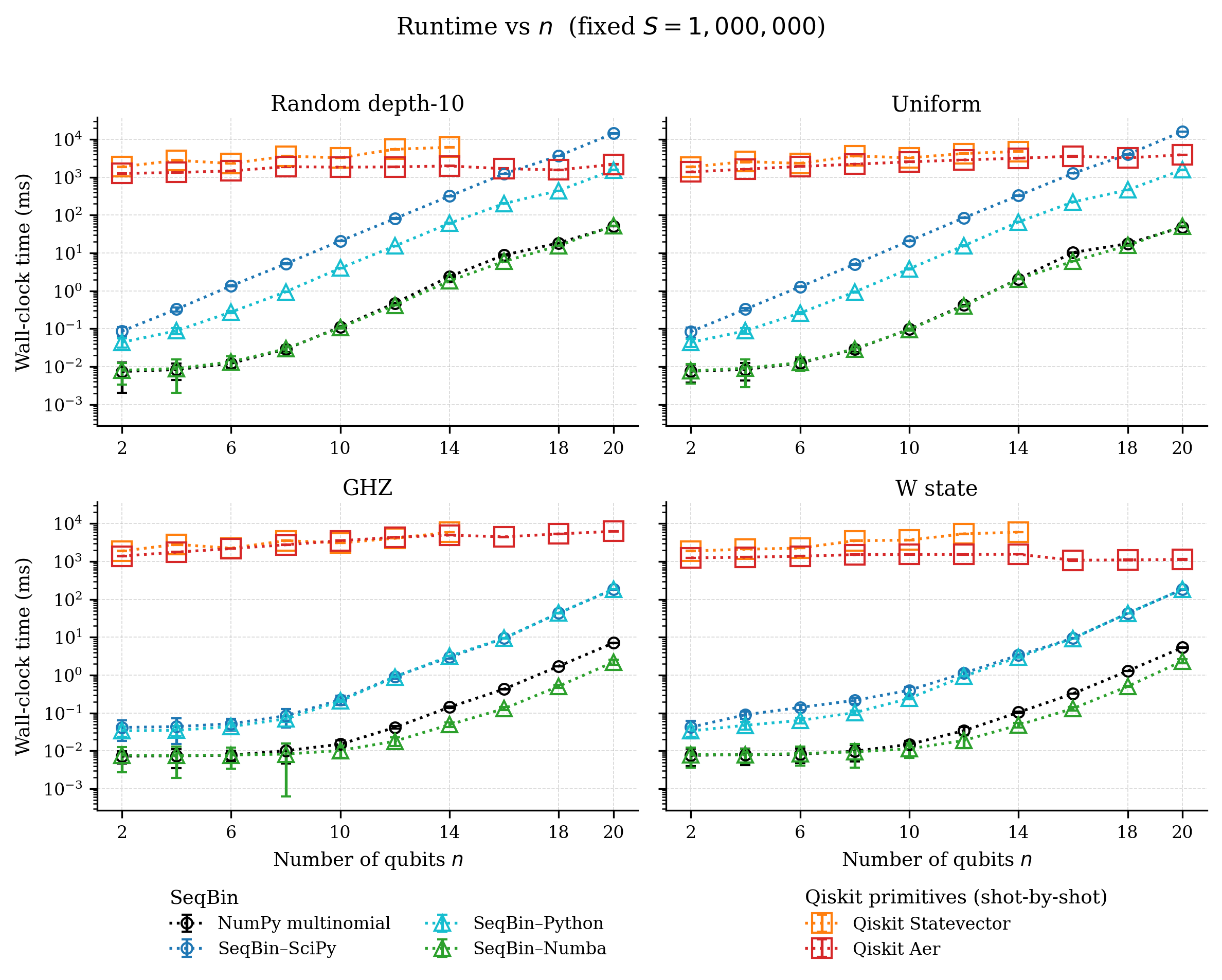}
    \caption{Wall-clock runtime as a function of the number of qubits $n$ at fixed shot budget $S = 10^6$. The count-level methods (NumPy multinomial and the three \seqbin{} backends) are timed over 50 repeats and the Qiskit primitives over 10, with two warmup calls each. Qiskit \texttt{StatevectorSampler} is omitted for $n \ge 16$ at this shot budget, due to memory issues.}
    \label{fig:runtime vs n}
\end{figure}

The previous section held the system size fixed and varied the shots. We now do the reverse. Figure~\ref{fig:runtime vs n} holds the shot budget at $S = 10^6$ and sweeps $n$ from 2 to 20, so the number of possible outcomes $D = 2^n$ grows from $4$ to about $10^6$. All four count-level methods follow the expected $\BigO{D} = \BigO{2^n}$ scaling, but with prefactors that vary across implementations.

On the dense states NumPy multinomial and \seqbin--Numba track each other closely, agreeing to within about ten per cent at most sizes, with \seqbin--Numba somewhat ahead between $n = 12$ and $18$; at $n = 20$ they reach about $50$ and $52$ ms respectively. On the sparse states \seqbin--Numba becomes faster as $n$ grows, because bunching confines its sweep to the outcomes that carry probability while NumPy still visits all $2^n$ of them: at $n = 20$ it is about three times faster on GHZ ($2.2$ against $7.1$ ms) and about twice as fast on W ($2.3$ against $5.4$ ms). \seqbin--Python is roughly $20$–$40\times$ slower once the sweep dominates, from $n = 6$ upward, because its inner loop runs in the interpreter; it is included as a pedagogical reference for the unadorned sequential decomposition rather than as a practical method.
The \seqbin--SciPy curve requires comment. It is the slowest of the \seqbin{} backends on the dense states, reaching the $10^4$~ms region by $n = 20$. The cause is not the algorithm but the location of the work: every outcome requires one call into SciPy's general-purpose distribution machinery, which performs argument checking and setup before generating a variate. At $n = 20$ there are approximately $10^{6}$ outcomes, so this fixed cost is incurred $10^{6}$ times. On the sparse states the same curve stays low, because bunching leaves only the few outcomes that carry probability ($2$ for GHZ, $n$ for W), and so only a handful of those calls are ever made.
 Qiskit Aer varies comparatively little with $n$, staying between about $1$ and $6$~s across the whole range, because at $S = 10^6$ its cost is dominated by the per-shot term. \texttt{StatevectorSampler} is comparable to Aer on the GHZ state and up to about four times slower on the other states for $n \leq 14$, and is omitted beyond that point because of its memory consumption.

\subsection{Effects of state structure}
\label{sec:state_structure}

\begin{figure}
    \centering
    \includegraphics[width=\linewidth]{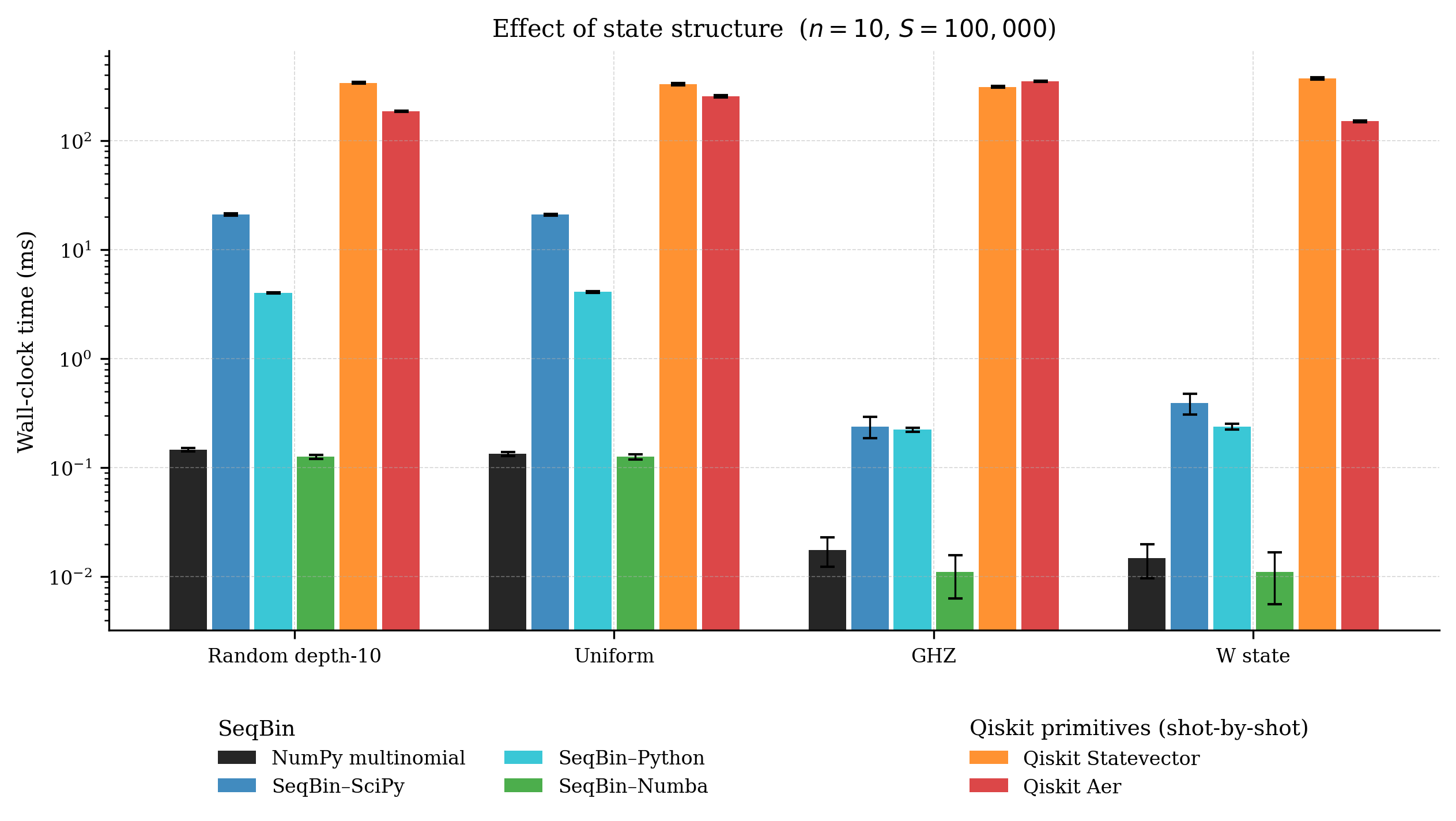}
    \caption{ Runtime at $n = 10$, $S = 10^5$, separated by state type (Random depth-10, Uniform, GHZ, W). Bars are the mean over 50 repeats for NumPy multinomial and the three \seqbin{} backends, and 10 repeats for the Qiskit primitives; whiskers show one standard deviation.}
    \label{fig:effect_of_state}
\end{figure}

In Figure~\ref{fig:effect_of_state} we fix both $n$ and $S$ and change only the state type. All four states have the same dimension $D = 2^n = 1024$, but their probability supports are very different. Random and Uniform are dense (every bin carries mass), while GHZ and W are sparse (only $2$ and $n = 10$ bins carry mass).

On the dense states, the ordering is set by per-bin overhead. NumPy multinomial and \seqbin--Numba are close to $0.1$ ms. \seqbin--Python is about $30\times$ slower and \seqbin--SciPy about $150\times$ slower; both pay a Python-level call cost on every bin. The Qiskit primitives are far slower again because their cost grows with $S$.

On the sparse states the absolute runtimes drop, and \seqbin--Numba is
faster than NumPy multinomial, at about $0.011$~ms against $0.015$ to
$0.018$~ms. NumPy touches all $1024$ bins, whereas the \seqbin{}
implementations collapse the zero-probability outcomes into a single rare bin
and sweep only the few that carry mass (Section~\ref{sec:bunching}). At this
small dimension the margin is modest, since a sweep over $1024$ bins is itself
cheap; it widens with $n$, as Figure~\ref{fig:runtime vs n} shows.
\seqbin--SciPy and \seqbin--Python benefit from the same shortcut, but their
per-bin overhead keeps them twenty to thirty-five times above \seqbin--Numba.
The Qiskit primitives gain nothing from the sparsity, since
both produce one outcome per shot however few outcomes carry probability.

\subsection{Checking that the samples are correct}
\label{sec:validation}

A reduction in runtime is only useful if the samples remain correct, since a
faster sampler that distorts the distribution is of no value. Before drawing any
conclusion from the runtimes above, we therefore verify that all six methods
produce the same statistics.

The comparison must allow for the fact that sampling is random. Two correct
samplers given the same probabilities will not return identical counts, and
neither will reproduce the exact probabilities perfectly. The question is
therefore whether each method comes as close to the exact distribution as random
sampling permits, and no further away.

We measure that distance directly. For each run the observed frequency of every
outcome, $k_i/S$, is compared with its exact probability $p_i$, and the
differences are summed:
\begin{equation}
\label{eq:l1}
L_1 = \sum_{i=1}^{D} \left| \frac{k_i}{S} - p_i \right| .
\end{equation}
This is the total amount by which the sampled distribution departs from the
exact one. It is zero for a perfect match and increases as the sample departs
further. A correct sampler does not drive $L_1$ to zero; it drives $L_1$ down to
the level set by the number of shots, and no lower.

\begin{figure}
    \centering
    \includegraphics[width=\linewidth]{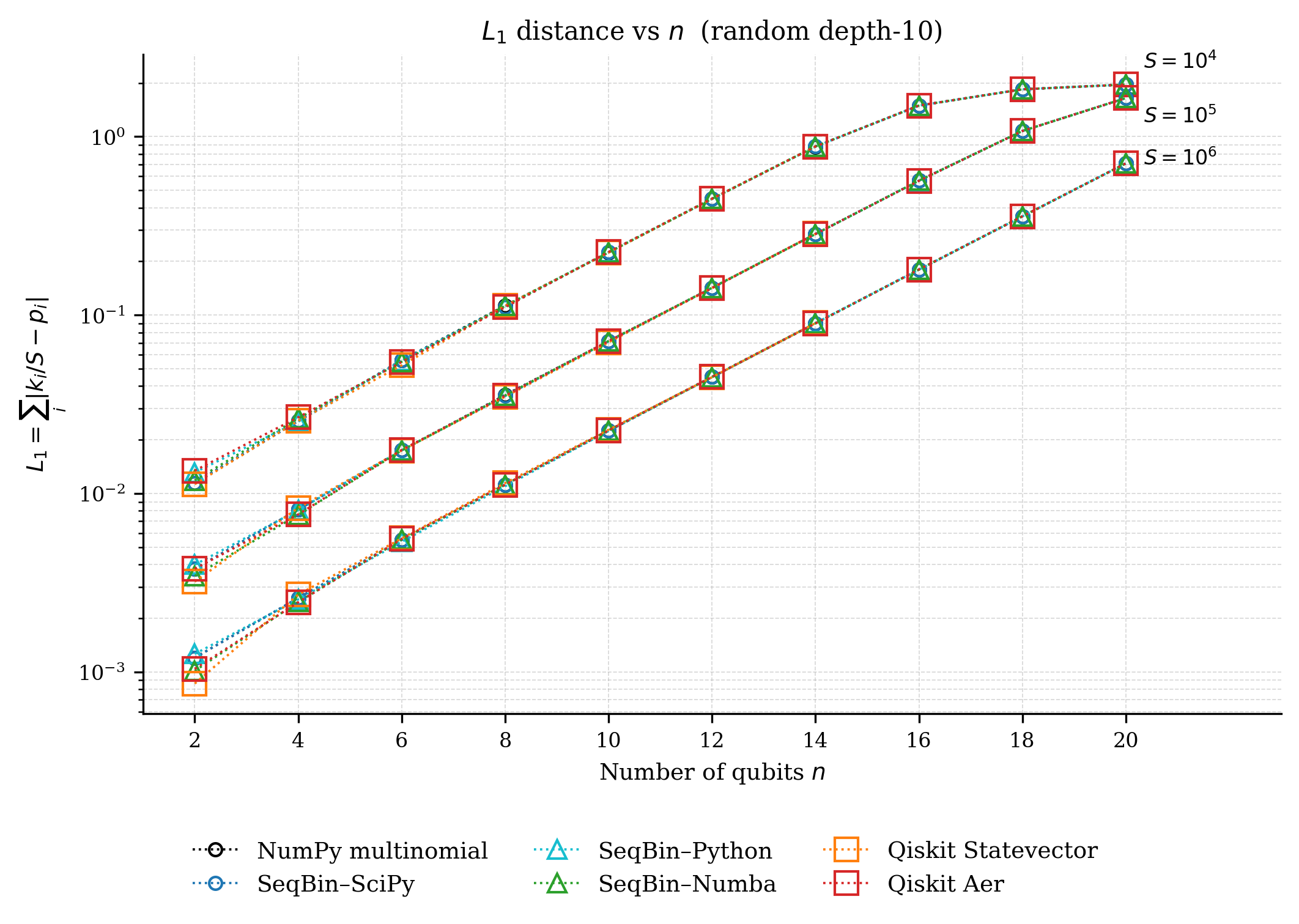}
    \caption{How far each sampler lands from the exact distribution, measured by $L_1$ of Eq.~(\ref{eq:l1}), against the number of qubits $n$, for the random depth-10 state. The three bands are three shot budgets: $S = 10^4$ (top), $10^5$, and $10^6$ (bottom). Lower is closer to exact. All six methods coincide in every band, so no method is closer to or further from the exact distribution than any other; the bands are separated by the shot budget rather than by the method. Markers follow the convention of the earlier figures: open circles (NumPy multinomial and \seqbin--SciPy), triangles (\seqbin--Python and \seqbin--Numba), squares (Qiskit). Each value is averaged over 50 independent draws, 10 for the Qiskit primitives.}
    \label{fig:l1}
\end{figure}

Figure~\ref{fig:l1} plots this quantity against the number of qubits, at three
shot budgets. Two features are relevant. First, every method coincides with
every other at every point, to within the width of a marker. There is no
separation between the \seqbin{} implementations and the library routines, or
between either of those and the Qiskit primitives. Second, the common curve behaves as sampling theory
says it should: the error shrinks as the shot budget grows, roughly as
$1/\sqrt{S}$, and it grows with $n$, because at a fixed number of shots the same
sample has to cover exponentially more outcomes.  At the largest $n$ the upper curves flatten just below $2$. This is a
ceiling imposed by the measure rather than a property of the samplers, and
the right-hand edge of the figure therefore carries no information about
their relative quality (\ref{app:l1ceiling}).

Taken together this says the six methods are statistically interchangeable: they
all reach the floor that the shot budget imposes, and none of them reaches it
from the wrong side. The differences reported in the preceding figures are
therefore differences in speed alone. We also ran a second, more standard test, a Pearson $\chi^2$ goodness-of-fit
test, which asks the same question in the form conventionally reported in the
literature. It agrees: every method is consistent with the exact distribution
across the whole sweep. Because that test needs some care in interpretation when
the distribution is sparse, we give it, and the full table of results, in
\ref{app:chi2}.

\section{Conclusion}
\label{sec:conclusion}

We have shown that the shot-sampling step of a classical quantum-circuit
simulation can be performed in expected $\BigO{2^n}$ time, independent of the
number of shots~$S$, by decomposing the multinomial draw into a sequence of
conditional binomial draws, each generated in expected constant time by the
BTPE algorithm. Rare-outcome bunching shortens the sweep further for structured
states. The resulting speedup over
Qiskit's current sampling primitives is
substantial in the regime $S \gtrsim 2^n$ in which measurement data is
collected, and the samples remain statistically indistinguishable from those of
the reference implementations.

The construction itself is classical, dating at least to
Devroye~\cite{Devroye1986}, with the BTPE primitive due to Kachitvichyanukul and
Schmeiser~\cite{BTPE1988} and its average-case behaviour analysed by
Cicirello~\cite{Cicirello2024}; the distributed variant was studied by
Webb~\cite{Webb2018}. The closest discussion within the quantum-computing
community is Qiskit issue~\#8618~\cite{QiskitIssue8618}. The contribution of
this work is to locate the per-shot bottleneck in the current reference
workflow, to provide an implementation that exploits the sparsity of
quantum-state probability distributions, and to validate the result across
system size, shot count and state structure. The pipeline profile in \ref{app:qiskit} shows that
statevector construction accounts for less than $2\%$ of the measured runtime in
both Qiskit paths, the remainder being sampling and shot-resolved result
handling.

The approach is most useful when $S$ is comparable to or larger than the number
of outcomes carrying appreciable probability. This is the regime of variational
algorithms such as the variational quantum eigensolver~\cite{Peruzzo2014} and
the quantum approximate optimization algorithm~\cite{Farhi2014} on near-term
systems, where $n \leq 20$ qubits and shot counts of $10^4$ to $10^6$ are
common. For larger systems, where reading the full probability vector is itself
the dominant cost, the focus should instead be on the simulation step. Where
only measurement counts are required, replacing the per-shot draw inside
\texttt{Statevector.sample\textunderscore counts} with a count-level draw would
preserve the Qiskit API and recover most of the speedup. A direct
\texttt{numpy.random.multinomial} call already achieves this; \seqbin--Numba
matches it on dense states and is faster on states whose probability is
concentrated on a few outcomes, by about a factor of three for a 20-qubit GHZ
state. Possible extensions include noisy simulation,
partial measurements and adaptive shot allocation.

\section*{Data availability statement}
The code and data that support the findings of this study are openly available
at the following URL: \url{https://github.com/ijaz-3/seqbin}.

\section*{Acknowledgments}
MP and IAM acknowledge support from the VEGA project No.~2/0055/23. This work
was supported by the use of computational resources of the supercomputer PERUN,
operated by the Supercomputing Centre at the Technical University of
Ko\v{s}ice (TUKE), Slovakia, with the support of the European Union from the
funds of the Recovery and Resilience Plan of the Slovak Republic within the
framework of project No.~17I03-04-P03-00001, Development and design of a
supercomputer for the National Supercomputing Center.

The authors used Claude (Anthropic, Claude Opus 5.5) for language editing and to
assist with parts of the benchmarking code; all text and code were checked by
the authors.


\appendix

\section{Where the Qiskit sampling cost goes}
\label{app:qiskit}
\label{sec:overhead}

\begin{figure}[h]
    \centering
    \includegraphics[width=\linewidth]{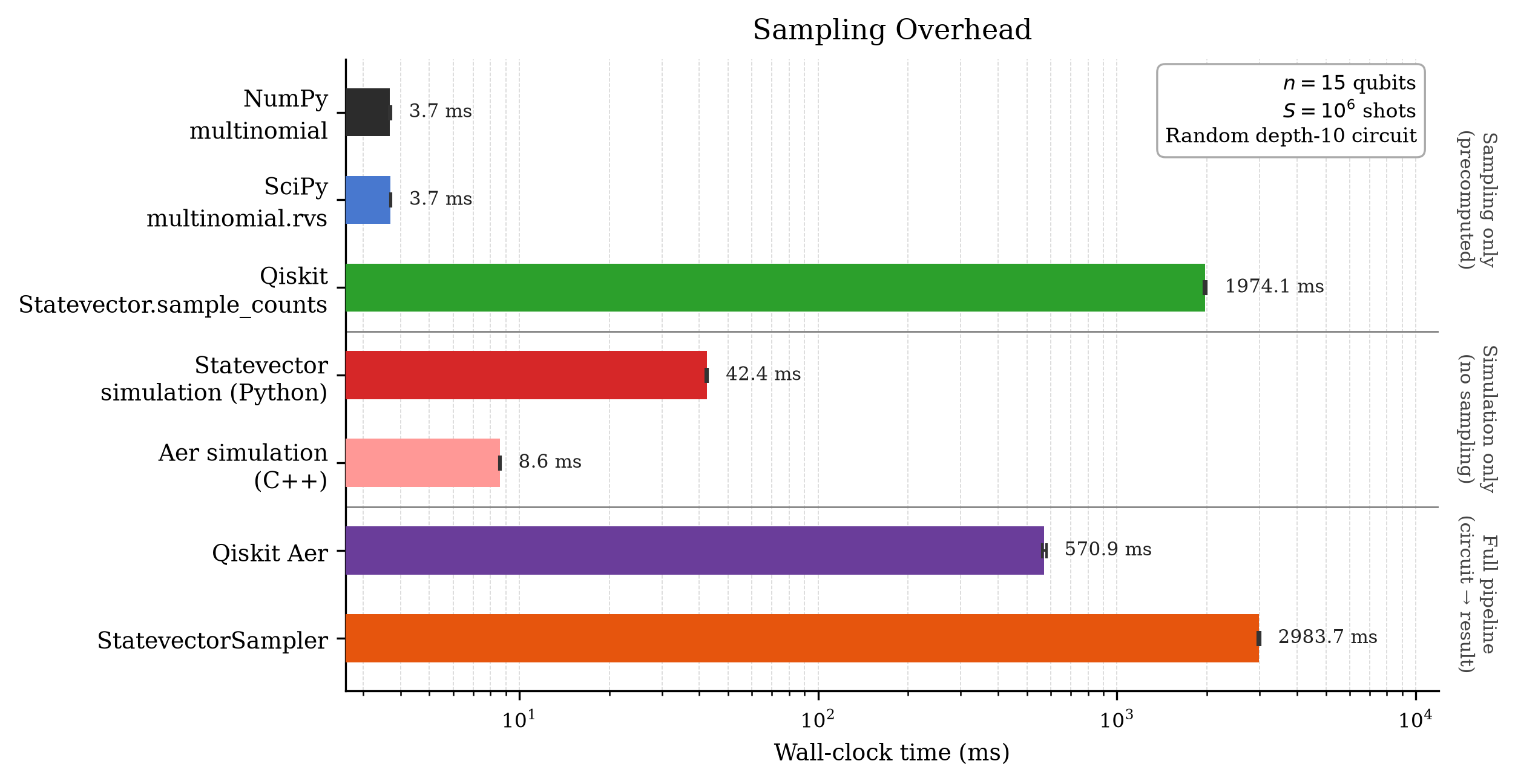}
    \caption{Decomposition of sampling cost at $n=15$ qubits, $S=10^6$ shots, random depth-10 circuit, on Perun (50 repeats; error bars are $\pm 1$ standard deviation). Sampling only: methods that draw $S$ shots from a precomputed distribution/statevector. Simulation only: time to build the statevector with no sampling, for the Python (Statevector) and C++ (AerSimulator) backends used internally by the two primitives. Full pipeline: end-to-end execution from the circuit to the shot-resolved result returned by each primitive, in the steady state.}
    \label{fig:overhead}
\end{figure}

Section~\ref{sec:problem} noted that Qiskit's reference sampler does work proportional to the number of shots. Here we locate that cost inside the pipeline. We benchmark a fixed workload of \(n=15\) qubits and \(S=10^6\) shots on a random depth-10 circuit (Figure~\ref{fig:overhead}).
The first comparison contrasts \texttt{numpy.random.multinomial} and \texttt{scipy.stats.multinomial.rvs}, both operating directly on a precomputed probability vector, with the full Qiskit sampler pipelines,  which accept a circuit and return a BitArray of per-shot outcomes (the V2 result object). The bare NumPy and SciPy multinomial calls both complete in approximately \(3.7\) ms.
In contrast, the full \texttt{StatevectorSampler} pipeline requires about \(2984\) ms, while the Aer sampler requires about \(571\) ms in the same steady-state setting.
The gap is far too large to be explained by sampling efficiency alone, but this two-way comparison does not yet localise where the cost originates inside the Qiskit pipeline.

We therefore extend the comparison with two additional classes of measurement. The first is \texttt{Statevector.sample\textunderscore counts} called on a pre-built \texttt{Statevector} object, which draws $S$ shots from a precomputed amplitude vector without re-simulating the circuit. This is the closest Qiskit-side analogue to \texttt{numpy.random.multinomial}, and we initially expected it to bound the cost of Qiskit's multinomial draw in isolation.

We can resolve this by looking into the Qiskit source~\cite{QiskitStatevectorSource}.\linebreak[0] \texttt{Statevector.\linebreak[0]sample\textunderscore memory} (the underlying call) generates each shot independently via \texttt{numpy.random.choice} over a length-$D$ array of bitstring labels, and \texttt{sample\textunderscore counts} then aggregates with \texttt{numpy.unique}. The cost is therefore $\BigO{S\log D + D}$ rather than the $\BigO{D}$ that \texttt{numpy.random.multinomial} would achieve. The choice is forced by the API contract, which exposes each shot as an individual bitstring rather than as a count vector.

We next separate the cost of producing the statevector from the cost of sampling and formatting the result.
For the Python statevector path used by \texttt{StatevectorSampler}, constructing the statevector from the circuit takes \(42.4\) ms. For the Aer C++ statevector path, the corresponding simulation-only
time is \(8.6\) ms.
These numbers are small compared with the full sampler pipelines: statevector construction accounts for only about \(1.4\%\) of the \texttt{StatevectorSampler} runtime and about \(1.5\%\) of the Aer runtime.
Thus, for this benchmark, the dominant cost is not the quantum-state simulation itself, but what happens after the statevector has been obtained: sampling, materialising the shot outputs, converting them into count data structures, and returning them through the
primitive interface.

\begin{figure}
    \centering
    \includegraphics[width=\linewidth]{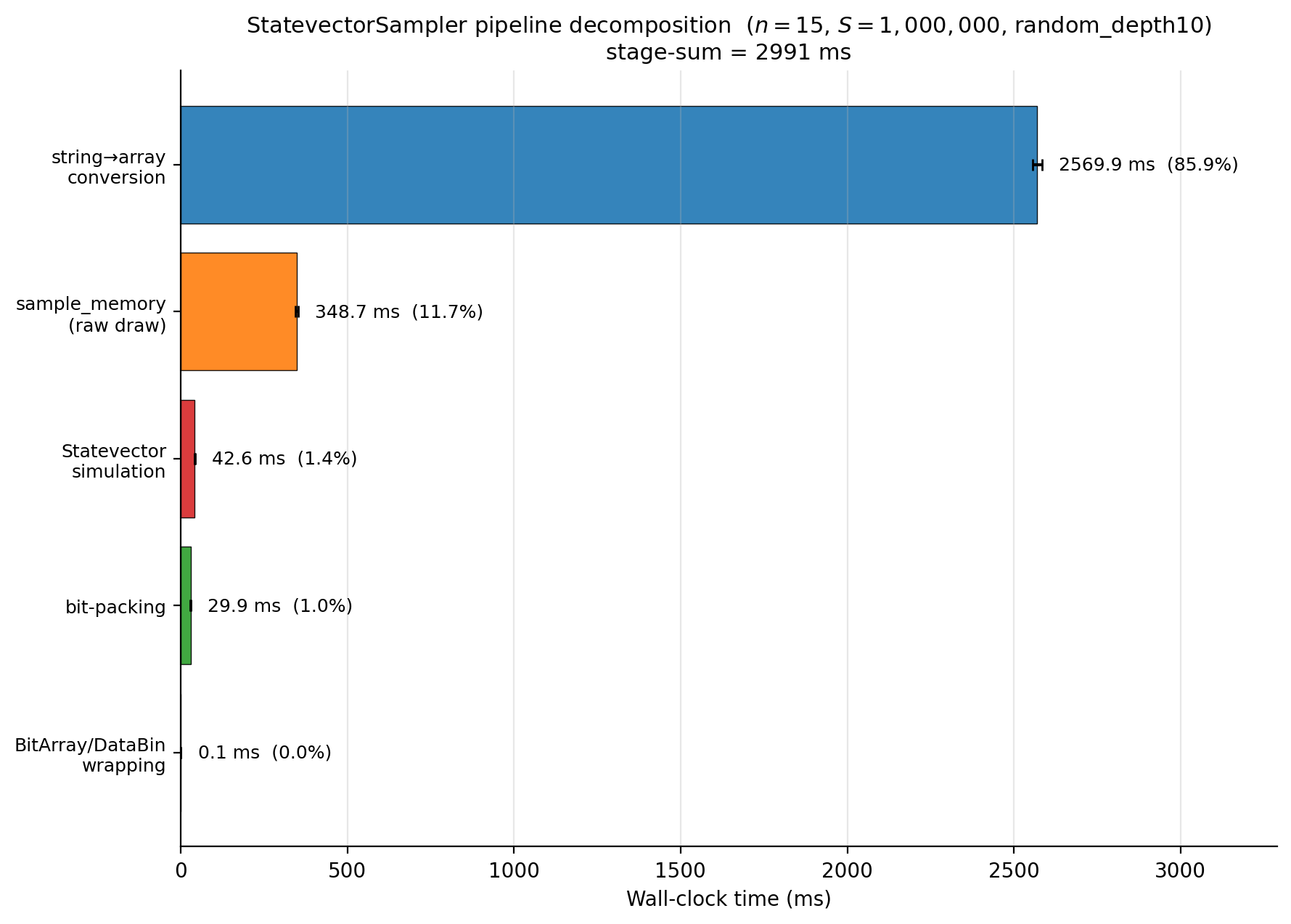}
    \caption{Detailed decomposition of the \texttt{StatevectorSampler} path for \(n=15\), \(S=10^6\), and a random depth-10 circuit. The measured stages sum to \(2991\) ms, consistent with the full \texttt{StatevectorSampler} runtime in Figure~\ref{fig:overhead}. Statevector simulation accounts for only \(42.6\) ms, while the raw \texttt{sample\_memory} draw takes \(348.7\) ms. The dominant contribution is the conversion of the sampled bitstrings into the array/count representation, which takes \(2569.9\) ms, or \(85.9\%\) of the decomposed pipeline. This shows that the main bottleneck is not statevector simulation, but the per-shot sampling and result-construction path.}
    \label{fig:statevector_pipeline_decomposition}
\end{figure}

To pinpoint which step inside that remainder dominates, we instrument \texttt{StatevectorSampler.\_run\_pub}~\cite{QiskitStatevectorSampler} stage by stage (Figure~\ref{fig:statevector_pipeline_decomposition}). Of the \(2991\) ms per-pub cost, the raw per-shot draw
(\texttt{sample\_memory}, internally \texttt{numpy.random.choice}) accounts for \(349\) ms (\(11.7\%\)), and the bit-packing into a \texttt{BitArray}
(\texttt{\_samples\_to\_packed\_array}) accounts for \(30\) ms (\(1.0\%\)).
The dominant stage, at \(2570\) ms or \(85.9\%\), is a single Python-level list comprehension,
\texttt{np.array([np.fromiter(s, uint8) for s in samples])},
that converts the \(S\) per-shot bitstring strings returned by \texttt{sample\_memory} into a numeric array. This is an
\(\mathcal{O}(S\cdot n)\) operation executed in interpreted Python.

Container wrapping (\texttt{BitArray}/\texttt{DataBin}) is negligible, at below \(0.1\%\), and primitive-framework dispatch is negligible too: the stage sum (\(2991\) ms) matches the full \texttt{run()} call (\(2984\) ms) to within
run-to-run noise.
We note that the V2 pipeline returns a \texttt{BitArray}, not a counts dictionary, so there is no counts-conversion step in the timed path. Thus, the bottleneck is not in the multinomial draw nor in the packing nor the framework. It is the materialisation and re-parsing of per-shot bitstrings, exactly
the work a count-vector representation avoids.

The cost is paid on every call rather than once at startup. In a freshly
started process, the module import (about $0.5$~s, loading the compiled backend)
is the only substantial one-time cost; the constructor takes under $0.1$~ms, and
the first call matches later calls to within timing noise. The runtimes reported
here are therefore steady-state per-call costs, which is the setting most
favourable to the primitives.

This decomposition settles the question. The roughly \(800\times\) gap between bare \texttt{numpy.random.multinomial} and \texttt{StatevectorSampler}, and the \(150\times\) gap against Aer, are dominated by framework-level per-shot handling rather than by differences in the underlying
sampling algorithm.
Any user who already has a probability distribution in hand,
whether from a single simulation, a noise-model evaluation, or a classical reference distribution, can recover the bulk of this speedup simply by calling \texttt{numpy.random.multinomial} directly.

For \texttt{StatevectorSampler}, the stage breakdown makes the split precise: the V2 algorithmic shift from \texttt{numpy.random.multinomial} to per-shot
\texttt{numpy.random.choice} accounts for only about an eighth of the cost, namely the \(11.7\%\) \texttt{sample\_memory} stage.
The larger part, roughly \(86\%\), is the framework-level per-shot formatting that materialises each draw as a bitstring and re-parses it into an array before packing.
The Aer pipeline incurs its overhead through a different, compiled sample-measure path, but the conclusion is the same:
the cost lies in producing and formatting per-shot outcomes, not in the multinomial draw itself.

The same pattern is found in PennyLane~\cite{PennyLane}. In its reference
simulator, \texttt{default.qubit}, a shot-based measurement first draws every
shot separately, with \texttt{numpy.random.Generator.choice} over the $2^n$
basis states~\cite{PennyLaneSource}. A \texttt{qml.counts} measurement then
converts each sampled shot into a bitstring in interpreted Python, joining its
bits one at a time, and aggregates the strings with \texttt{numpy.unique}. The
compiled \texttt{lightning.qubit} simulator generates the shots in
C\texttt{++}~\cite{LightningSource}, but returns them as individual samples that
pass through the same Python counting step. In both cases the sampling cost
grows with $S$, as it does for the Qiskit primitives. We confirmed this
with a short test on a laptop, separate from the Perun benchmarks of
Section~\ref{sec:comparison}, using PennyLane~0.45.1 and
PennyLane-Lightning~0.45.0. For a
random $15$-qubit state measured with \texttt{qml.counts}, the runtime of both
simulators grows in proportion to $S$, from a few tens of milliseconds at
$S = 10^4$ to several seconds at $S = 10^6$, about a thousand times slower than
a direct \texttt{numpy.random.multinomial} draw on the same probabilities. Of
the sampling work in \texttt{default.qubit}, the counting step takes about
$95\%$ and the per-shot draw about $5\%$. As in Qiskit, the cost lies in formatting the individual shots rather
than in drawing them.

\section{Technical details of BTPE}
\label{app:btpe}

The acceptance test in the BTPE sampler requires the binomial probability
mass at the proposed integer $k$,
\begin{equation}
\label{eq:logprob}
P(X=k) = \binom{N}{k}p^k(1-p)^{N-k}.
\end{equation}
For large $N$, this expression is evaluated in log-space:
\begin{eqnarray}
\log P(X = k) &= \log\Gamma(N+1) - \log\Gamma(k+1)
              - \log\Gamma(N-k+1) \nonumber\\
              &\quad + k\log p + (N-k)\log(1-p),
\end{eqnarray}
where $\Gamma(n)=(n-1)!$ for positive integers. Standard numerical libraries
provide fast implementations of $\log\Gamma$, such as \texttt{math.lgamma} in
Python and \texttt{lgamma} in C. This makes the verification step a constant
number of special-function evaluations.

BTPE is used when the expected number of successes is sufficiently large,
typically $Np\geq 30$ and, by symmetry, $N(1-p)\geq 30$. When either expected
count is smaller, an inverse-CDF or another small-mean method is used instead.
In the sequential decomposition this situation occurs for outcomes with small
conditional probability, or when only a small number of shots remain.

\section{Implementation details and low-occupancy safeguards}
\label{app:implementation}

This appendix collects the parts of the implementation that are needed to
reproduce the benchmarks but that do not carry the main result: how the
remaining probability is tracked, how the rare-outcome threshold is kept from
misbehaving when shots are scarce, when the sweep stops or hands its tail
to a categorical draw, and why the outcomes are not sorted.

\subsection{Tracking the remaining probability by suffix sums}
\label{app:suffix}

At step $i$ of the sweep, Algorithm~\ref{alg:seqbinom} needs the total probability
still unassigned,
\begin{equation}
\label{eq:suffix}
r_i = \sum_{j=i}^{D} p_j ,
\end{equation}
so that the conditional probability for outcome $i$ is $p_i / r_i$. The
straightforward way to obtain it is to keep a running total and subtract each
$p_i$ as it is used, which is what line~9 of Algorithm~\ref{alg:seqbinom}
describes.

We instead compute all $D$ values of $r_i$ once before the sweep starts, working
backwards from the last outcome: $r_D = p_D$, then $r_{D-1} = p_{D-1} + r_D$, and
so on. This backward pass touches each probability once and therefore costs
$\BigO{D}$, the same as the sampling sweep it accompanies, so it does not change
the complexity of the method.

The reason to prefer it is arithmetic rather than speed. Every subtraction in a
running total introduces a small rounding error, and those errors accumulate
along the sweep. They matter most at the end, where $p_\mathrm{rem}$ has become
small and the ratio $p_i / p_\mathrm{rem}$ is most sensitive to an error in its
denominator. Computing each $r_i$ directly from the probabilities avoids passing
an error forward from one step to the next. Because every $p_j$ is
non-negative, the $r_i$ never increase along the sweep, whatever the order in
which the outcomes are visited.

We did not observe any difference in the statistical tests of
Section~\ref{sec:validation} at the shot counts and precision benchmarked here.
We use the suffix sums because they cost the same as the running total, not
because the running total was measurably failing.

\subsection{The adaptive rare-outcome threshold}

For the benchmarks we use the specific choice \(c=c_{\rm rare}=1\), so that an outcome is classified as rare when its expected count satisfies \(S p_i \lesssim 1\).
This choice is a heuristic, not a theoretically sharp threshold. It simply captures the intuition that outcomes expected to receive less than one shot do not need to be treated individually in the first pass.

The fixed threshold \(\theta=c_{\rm rare}/S\) assumes that the shot budget is large
enough relative to the number of outcomes. When \(S \lesssim D\), this rule can
backfire. For example, in the uniform case \(p_i=1/D\), all outcomes are classified
as rare whenever \(S \lesssim D\). The rare bucket then contains essentially all shots,
and redistributing it reduces to categorical sampling over the full outcome space, with
cost \(\BigO{S\log D}\), which can be worse than the plain \(\BigO{D}\) \seqbin{}
sweep.

To avoid this failure mode, we use a conservative adaptive threshold. Let
\(r=S/D\) be the average number of shots per outcome. We replace \(c_{\rm rare}\) by
\[
\tilde c_{\rm rare}=c_{\rm rare}\min(1,r/10).
\]
Thus, for \(r>10\) the usual rare-outcome rule is unchanged, while for \(r\leq 10\)
the threshold is reduced in proportion to the average occupancy. The numerical factor
10 is a heuristic safeguard chosen to keep rare-outcome bunching inactive in the
low-shot-density regime where it can be counterproductive. In addition, if the adaptive
rule would classify all outcomes as rare, we bypass bunching and run the plain
\seqbin{} sweep instead. This fallback prevents the main pathological case
in which the rare bucket would replace an \(\BigO{D}\) sweep by an \(\BigO{S\log D}\)
categorical redistribution step.

\subsection{Early stopping}
\label{sec:earlystop}
Once $S_\mathrm{rem}$ reaches zero every shot has been assigned, so all
remaining counts are zero and the sweep can stop. This is a one-line check and
it costs nothing to include, but it is worth being clear about how much it
actually does. It only fires when the shots run out before the outcomes do,
which requires a long tail of outcomes still to be visited. On the sparse states
we benchmark, the loop has already been shortened by the bunching of
Section~\ref{sec:bunching} before this check can matter: for a GHZ state at
$n=10$, the $2$ outcomes carrying probability form the common set and the
remaining $1022$, which carry no probability, are dropped, so the sweep is one
iteration long to begin with. Early stopping is therefore best understood as a
safeguard for the low-shot regime, $S \lesssim D$, rather than as the mechanism
behind the sparse-state runtimes reported in
Section~\ref{sec:state_structure}.

NumPy's multinomial generator~\cite{NumpySource} also stops once the remaining
shot count reaches zero, and returns immediately for outcomes of zero
probability, so the benefit of early stopping is not specific to our
implementation. What distinguishes the implementations compared here is the
per-call overhead of preparing the probability vector and constructing the
dense count output.

\subsection{Switching to a categorical tail}
\label{sec:switching}
When $S_\mathrm{rem}$ drops below a threshold ($S_\mathrm{rem} < 100$ in our
implementation),
it may be faster to switch to direct categorical sampling for the remaining
shots, since each categorical draw costs $\BigO{\log D} = \BigO{n}$ via
binary search on the (already available) CDF, and the total cost
$\BigO{S_\mathrm{rem} \cdot n}$ is small.

\subsection{Order of the outcomes}
\label{app:sortcost}
The decomposition is exact for any order in which the outcomes are visited,
and the sweep takes them in their natural index order. Sorting them by
decreasing probability is a common refinement, intended to draw the large
outcomes while $S_\mathrm{rem}$ is still large. We do not use it here. Sorting
$D$ numbers costs $\BigO{D \log D}$, which exceeds the $\BigO{D}$ cost of the
sweep, and it requires the probabilities to be permuted before the sweep and
the counts to be permuted back afterwards.

To measure what sorting would cost, both variants
were run with the \seqbin--Numba backend over the full benchmark grid of
Section~\ref{sec:bench_setup}, in the same jobs, so the comparison is free of
differences between runs. The median ratio of sorted to unsorted runtime lies
between $1.2$ and $1.7$ for $n \leq 8$, and between $1.1$ and $1.2$ on the
dense states for $n \geq 10$. On the sparse states it is $3.2$, $9.6$ and $15$
at $n = 10$, $14$ and $20$, where bunching has reduced the sweep to a few
outcomes and the sort is left as the dominant cost. The unsorted sweep was
never measurably slower than the sorted one. Without the sort, outcomes of zero probability may
precede positive ones, so the categorical tail of \ref{sec:switching} uses a
strict comparison against the cumulative distribution; this ensures that no
shot is assigned to an outcome of zero probability.

\section{Further statistical checks}
\label{app:chi2}

\subsection{The ceiling of the $L_1$ distance}
\label{app:l1ceiling}
The $L_1$ distance of Eq.~(\ref{eq:l1}) is bounded above by $2$. Since $k_i/S$
and $p_i$ are both non-negative and both sum to one,
$|k_i/S - p_i| \leq k_i/S + p_i$ term by term, and summing gives
$L_1 \leq 1 + 1 = 2$. The bound is attained only when the sampled and exact
distributions have disjoint support. Deep undersampling approaches it: at
$n = 20$ with $S = 10^4$ there are about one hundred possible outcomes per
shot, so almost every outcome is unobserved. The outcomes that were observed
carry almost none of the true probability, contributing approximately one to
the sum, and the outcomes carrying the probability were not observed,
contributing approximately one more. The flattening of the upper curves in
Figure~\ref{fig:l1} therefore records exhaustion of the shot budget rather than
degradation of the sampler.

\subsection{Goodness-of-fit test}

Section~\ref{sec:validation} compares the sampled and exact distributions by
their total absolute difference. Here we report the conventional alternative, a
Pearson $\chi^2$ goodness-of-fit test.

For a sample of $S$ shots drawn from $\mathbf{p}$, the statistic is
\begin{equation}
\chi^2 = \sum_{i=1}^{D} \frac{(k_i - S p_i)^2}{S p_i},
\end{equation}
which compares each observed count with the count expected under $\mathbf{p}$,
weighting each discrepancy by how large a fluctuation is normal for that bin.
Dividing by the number of degrees of freedom, $\mathrm{dof} = D - 1$, gives a
quantity whose mean is one for a correct sampler, with standard deviation
$\sqrt{2/\mathrm{dof}}$. Values near one indicate that the counts deviate from
$\mathbf{p}$ by about as much as chance alone would produce; values far from one
in either direction indicate a sampler that is wrong, or a fluke.

Table~\ref{tab:validation} reports $\chi^2/\mathrm{dof}$ for the random depth-10
state at each $(n, S)$. Every entry is consistent with unity. The spread is
wider at small $n$, which is expected rather than suspicious: with
$\mathrm{dof} = 3$ at $n = 2$, the $95\%$ interval for $\chi^2/\mathrm{dof}$ is
roughly $[0.07, 3.12]$, so scatter of that size carries no information. The
Qiskit columns scatter more than the others throughout because they are averaged
over 10 draws rather than 50. Where the \seqbin--SciPy and NumPy entries
coincide exactly, the two methods have drawn the same binomial variates from
the same seeded generator in the same order, and so returned identical samples.

We report a dense state because $\chi^2/\mathrm{dof}$ is most informative when
the support is large. On sparse states the degrees of freedom collapse: the GHZ
state has only two outcomes with nonzero probability, so $\mathrm{dof} = 1$ and
the statistic scatters by order unity no matter how correct the sampler is. This
is precisely the case the $L_1$ measure of Section~\ref{sec:validation} handles
without difficulty, which is why we lead with it there.

\begin{table}[!ht]
\centering
\caption{Statistical validation on the random depth-10 (dense) state:
$\chi^2/\mathrm{dof}$ for every method at each $(n, S)$. A correct sampler
gives $\chi^2/\mathrm{dof}$ with mean unity and standard deviation
$\sqrt{2/\mathrm{dof}}$; for this dense state $\mathrm{dof} = 2^n-1$ is large,
so the expected scatter is small and every entry is consistent with unity. The wider spread at small $n$ reflects the few
degrees of freedom there ($\mathrm{dof}=3$ at $n=2$), and the larger scatter of
the Qiskit columns reflects their smaller sample (10 draws versus 50). Each
value is averaged over 50 independent multinomial draws for the
count-level methods and 10 for the Qiskit primitives. Cells marked `---'
are skipped runs: Qiskit \texttt{StatevectorSampler} is not run for $n \ge 16$,
at any shot count, because its per-shot bitstring formatting exceeds the memory
budget at $S=10^6$.
Cross-method agreement on all states, including the sparse ones, is shown by the
$L_1$ distance in Fig.~\ref{fig:l1}.}
\label{tab:validation}
\small
\setlength{\tabcolsep}{4pt}
\begin{tabular}{rr cccccc}
\toprule
$n$ & $S$ & NumPy & \seqbin & \seqbin & \seqbin & Qiskit & Qiskit \\
    &     & mult. & SciPy & Python & Numba & SV     & Aer \\
\midrule
2  & $10^4$ & 1.052 & 1.052 & 1.163 & 1.090 & 1.155 & 1.217 \\
2  & $10^5$ & 1.084 & 1.084 & 1.082 & 0.945 & 0.820 & 1.387 \\
2  & $10^6$ & 1.081 & 1.081 & 1.076 & 0.882 & 0.764 & 0.779 \\
\addlinespace
4  & $10^4$ & 0.926 & 0.926 & 0.935 & 1.019 & 1.013 & 1.140 \\
4  & $10^5$ & 0.952 & 0.952 & 0.971 & 0.917 & 1.067 & 0.945 \\
4  & $10^6$ & 0.980 & 0.980 & 0.975 & 0.939 & 0.992 & 0.864 \\
\addlinespace
6  & $10^4$ & 1.039 & 1.039 & 1.025 & 1.019 & 0.965 & 0.984 \\
6  & $10^5$ & 1.015 & 1.015 & 1.016 & 0.970 & 1.030 & 1.029 \\
6  & $10^6$ & 1.009 & 1.009 & 0.968 & 1.005 & 1.046 & 0.959 \\
\addlinespace
8  & $10^4$ & 1.007 & 0.984 & 1.002 & 1.008 & 0.998 & 0.978 \\
8  & $10^5$ & 1.023 & 1.022 & 0.992 & 0.993 & 1.006 & 0.984 \\
8  & $10^6$ & 1.008 & 1.008 & 0.964 & 1.001 & 1.043 & 1.008 \\
\addlinespace
10 & $10^4$ & 1.000 & 1.009 & 1.000 & 0.987 & 0.999 & 1.025 \\
10 & $10^5$ & 0.996 & 0.999 & 1.000 & 1.008 & 0.987 & 1.012 \\
10 & $10^6$ & 1.000 & 1.000 & 0.993 & 1.009 & 1.020 & 0.991 \\
\addlinespace
12 & $10^4$ & 0.995 & 0.998 & 1.001 & 0.997 & 0.993 & 0.991 \\
12 & $10^5$ & 0.995 & 0.997 & 0.998 & 1.000 & 0.995 & 0.993 \\
12 & $10^6$ & 1.005 & 1.002 & 0.998 & 0.999 & 0.993 & 0.999 \\
\addlinespace
14 & $10^4$ & 0.997 & 1.007 & 0.997 & 1.002 & 1.000 & 1.010 \\
14 & $10^5$ & 0.999 & 0.999 & 0.995 & 1.002 & 0.999 & 0.997 \\
14 & $10^6$ & 0.996 & 0.999 & 1.000 & 0.999 & 1.003 & 0.995 \\
\addlinespace
16 & $10^4$ & 1.001 & 1.013 & 0.991 & 0.998 & ---   & 0.998 \\
16 & $10^5$ & 1.001 & 1.000 & 0.997 & 1.000 & ---   & 0.995 \\
16 & $10^6$ & 1.000 & 1.000 & 0.998 & 0.997 & ---   & 1.000 \\
\addlinespace
18 & $10^4$ & 0.996 & 1.001 & 0.995 & 0.997 & ---   & 0.994 \\
18 & $10^5$ & 1.000 & 1.001 & 0.999 & 1.000 & ---   & 0.998 \\
18 & $10^6$ & 1.000 & 1.000 & 0.999 & 1.000 & ---   & 1.000 \\
\addlinespace
20 & $10^4$ & 1.004 & 0.995 & 0.998 & 1.000 & ---   & 0.996 \\
20 & $10^5$ & 1.003 & 0.999 & 1.000 & 1.000 & ---   & 0.999 \\
20 & $10^6$ & 1.000 & 0.999 & 1.000 & 1.000 & ---   & 0.999 \\
\bottomrule
\end{tabular}
\end{table}

\section{An alternative algorithm: prefix-uniform sampling}
\label{app:prefix}

For completeness, we describe an alternative algorithm for sampling a
single binomial variate that does not rely on BTPE. While it is superseded
by BTPE for the application in this paper, it may be of independent interest
as a memory-efficient approach that avoids precomputing the full CDF.

The probability of obtaining exactly $k$~zeros in $N$~throws of a biased
coin with $P(0) = p$ is given by the binomial distribution
\begin{equation}
\label{eq:binomial}
b(k; N, p) = \binom{N}{k} p^k (1-p)^{N-k}.
\end{equation}
These bin probabilities can be computed recursively:
\begin{equation}
\label{eq:recurrence}
b(k+1; N, p) = b(k; N, p) \cdot \frac{N-k}{k+1} \cdot \frac{p}{1-p}.
\end{equation}
For large $N$, working in log-space avoids overflow:
\begin{equation}
\label{eq:logrecurrence}
\log b(k+1) = \log b(k) + \log(N-k) - \log(k+1) + \log p - \log(1-p),
\end{equation}
starting from $\log b(0) = N \log(1-p)$.

The prefix-uniform algorithm proceeds as follows:
\begin{enumerate}
    \item Start at the mode $L = \mathrm{round}(Np)$.
    \item Expand outward alternately: $L, L{-}1, L{+}1, L{-}2, L{+}2, \ldots$,
    computing bin probabilities recursively.
    \item Generate a random binary fraction $q = 0.b_1 b_2 b_3 \ldots$ bit-by-bit.
    \item After each new bit, check if the interval $[q, q + \varepsilon)$
    lies entirely within one bin's cumulative slice. If so, return that bin index.
    \item Otherwise, generate another bit and (if necessary) expand to more bins.
\end{enumerate}
This approach requires only two active probability values and cumulative
totals, with no precomputation of the full distribution. In the typical case
(landing near the mode), only a small number of bins need to be evaluated.


\begin{thebibliography}{99}
\bibitem{NoteQiskitDiscussion}
Note. The Qiskit discussion in~\cite{QiskitIssue8618} compares per-shot
categorical sampling against a single multinomial count draw and reports that
the count-level route becomes preferable well before the shot count reaches
the outcome dimension. The legacy V1 \texttt{Sampler} did call
\texttt{numpy.random.multinomial}; it was removed in Qiskit~2.0 and replaced by
the V2 \texttt{StatevectorSampler}, which draws each shot separately with
\texttt{numpy.random.choice}.
\bibitem{NoteLowShotRegime}
Note. For completeness: if the shot count were driven far below the size of
the support, so that most outcomes are never observed, a shot-by-shot draw
would eventually become competitive, since it need not process outcomes that
receive no shots. Such a run is too sparsely sampled to characterise the
output distribution and is not a regime in which measurement data is
collected, so we do not pursue it.
\end{thebibliography}
\end{document}